\documentclass[aps,pra,twocolumn,showpacs]{revtex4}
\usepackage{amssymb}
\usepackage{amsmath}
\usepackage{graphicx}
\usepackage{epstopdf}
\usepackage{bm}
\usepackage{braket}
\usepackage{color}
\usepackage{rotating,booktabs}

\begin{document}
\title{One-Body Density Matrix and Momentum Distribution of Strongly Interacting One-Dimensional Spinor Quantum Gases}

\author{Li Yang$^1$, and Han Pu$^{1,2}$}

\affiliation{$^{1}$Department of Physics and Astronomy, and Rice Center for Quantum Materials,
	Rice University, Houston, TX 77251, USA \\
	$^2$Center for Cold Atom Physics, Chinese Academy of Sciences, Wuhan 430071, P. R. China}


\begin{abstract}
The one-body density matrix (OBDM) and the momentum distribution of quantum many-body systems are usually very difficult to calculate. Here we develop a technique to calculate the OBDM and the momentum distribution of a general one dimensional (1D) spinor quantum gas in the strong interaction regime. This technique relies on a remarkable connection between the OBDM of the spinor gas and that of a spinless 1D hard-core anyon gas, which allows us to efficiently calculate the OBDM of the spinor system with particle numbers much larger than what was previously possible. Given the OBDM, we can easily calculate the momentum distribution of the spinor system, which is also related to the momentum distribution of the hard-core anyon gas. Our study not only provides a practical method for the calculation of the OBDM, but also provides significant new insights into the properties of 1D strongly interacting spinor quantum gases.
\end{abstract} 


\maketitle

\section{Introduction}

One dimensional (1D) quantum many-body systems possess many remarkable properties and have fascinated theorists for many decades \cite{book}. With the advent of cold atoms, we now have a experimentally realizable 1D system amenable to exquisite control \cite{Guan2013}. Recently, physics of a 1D spinor quantum gas in the strong interaction regime has been studied by using a strong coupling ansatz wave function \cite{Deuretzbacher2008,Guan2009,Deuretzbacher2014,Yang2015}, according to which, the many-body wave function of the system \cite{spinchargeproduct,Yang2016} can be mapped to a direct product of a spatial wave function described by a spinless fermion and a spin wave function governed by an effective spin-chain Hamiltonian first proposed in \cite{Deuretzbacher2014}. Over the past couple of years, there have been numerous works on this spin-chain model for strongly interacting pure spinor quantum gases \cite{Volosniev2014,Levinsen2015,Massignan2015,YangLJ2016,Volosniev2016,Loft2016_1,Loft2016_2} or Bose-Fermi mixtures \cite{Hu2016,Bellotti2016,Deuretzbacher2016_bfm}, from ground state properties to dynamics. A recent experiment investigated a few-body spin-1/2 Fermi gas in the strongly interacting regime \cite{Murmann2015}.

It is well known that, even with the knowledge of the many-body wave function, the calculation of correlation functions and momentum distribution of any quantum many-body system is in general extremely difficult and poses a tremendous challenge. This difficulty stems from the intrinsic complexity of the many-body wave function. The goal of the current paper is to propose a very efficient method of calculating the one-body density matrix and the momentum distribution of a strongly interacting 1D spinor gas by exploiting a remarkable connection between the OBDM of such a spinor gas and the OBDM of a spinless hard-core anyon gas. With this method, we can readily calculate the momentum distribution of a 1D quantum gases up to a few hundred particles, which is an order of magnitude larger than what was previously possible. Furthermore, this method also provides significant new insights into the 1D strongly interacting regime. For example, we show that the momentum distribution of a single impurity moving in a background of strongly interacting spinless bosons, which was measured in a recent experiment \cite{Meinert2016}, mimics that of a hard-core spinless anyon with a time-dependent statistical parameter. 


\section{One-Body Density Matrix}

Consider a spinor quantum gas with $N$ atoms. The explicit form for a strong coupling ansatz wave function with a single spatial wave function $\varphi$ (often referred to as the charge state in literature, which describes the particles distribution in position space) can be written as
\begin{equation}
\Psi(x_1,...,x_N;\sigma_1,...,\sigma_N) = \sum_P(\pm1)^PP(\varphi\theta^1\chi)\,, \label{SCAW}
\end{equation}
where $x_i$ and $\sigma_i$ denote the spatial and spin coordinates, respectively; $\pm1$ are for bosonic and fermionic gases, respectively; $P$ are permutation operators acting on both the spatial and the spin coordinates; $\varphi (x_1,...,x_N)$ is a spinless fermion wave function, $\theta^1$ is the generalized step function which restricts the system to the spatial sector $x_1<x_2<...<x_N$; and finally, $\chi (\sigma_1,...,\sigma_N)$ is a spin wave function for an $N$ sites spin chain system governed by the spin-chain Hamiltonian which takes the following form:
\begin{equation}
H_{\text{sc}}=-\sum_{j=1}^{N-1}C_j\frac{1\pm{\cal E}_{j,j+1}}{g}\,, \label{Hsc}
\end{equation}
where the coupling coefficients $C_j$ depend only on charge state $\varphi$, and ${\cal E}_{j,j+1}$ is the spin exchange operator that exchanges two neighboring spins \cite{Yang2015}; $g$ characterizes the interaction strength. For a system with spin-independent interaction, $g$ is a single number; in general, $g$ can also be an operator that has different values in different spin channels. The wave function represented by Eq.~(\ref{SCAW}) can be understood as having $N$ fermions with distribution probability amplitude given by $\varphi$, and with each fermion attached with a spin, which may be regarded as the continuum version of a slave fermion state. The corresponding one-body density matrix (OBDM) associated with the many-body wave function $\Psi$ is defined as
\begin{widetext}
\begin{equation}
\rho(x',x;\sigma',\sigma)
=\sum_{\sigma_1,...,\sigma_{N-1}}\int dx_1... dx_{N-1} \,\Psi^*(x_1,..., x_{N-1}, x';\sigma_1,...,\sigma_{N-1},\sigma')\Psi(x_1,..., x_{N-1},x;\sigma_1,...,\sigma_{N-1},\sigma)\,. \label{OBDM}
\end{equation}
Substituting Eq.~(\ref{SCAW}) into Eq.~(\ref{OBDM}), we have 
\begin{equation}
\rho(x',x;\sigma',\sigma)=\sum_{\sigma_1\cdots\sigma_{N-1}}\int dx_1\cdots dx_{N-1}\varphi'^*\varphi\sum_{P'P}\theta'^{P'}\theta^{P}\otimes(P'\chi'^{\dagger})(P\chi)\,,\label{obdm1}
\end{equation}
where we have used the short-hand notation $\varphi' = \varphi(x_1,..., x_{N-1},x')$, $\varphi = \varphi(x_1,..., x_{N-1},x)$, $\chi' = \chi(\sigma_1,...,\sigma_{N-1},\sigma')$, and $\chi = \chi(\sigma_1,...,\sigma_{N-1},\sigma)$. To evaluate the above equation, we need to order $x'$ and $x$ with respect to $x_1, \,\dots,\, x_{N-1}$. For example, assuming $x'<x$, we can take $x' \in (x_{m-1},x_m)$ and $x \in (x_{n-1},x_n)$ with $m\le n$, and denote this ordering configuration as $\Gamma_{m,n}$, in which 
\begin{equation}
\Gamma_{m,n}:\;\;x_{1}<...<x_{m-1}<x'<x_m<...<x_{n-1}<x<x_n<...<x_{N-1}\,.
\end{equation}
\end{widetext}
Once the ordering of $x'$ and $x$ are fixed, all permutations on $1\cdots N-1$ will lead to the same integral value, because these kind of permutations does not change either $\theta'^{P'}\theta^{P}$ or $(P'\chi'^{\dagger})(P\chi)$ . According to this observation, the OBDM~(\ref{obdm1}) can be written as \cite{Yang2015,Deuretzbacher2016}
\begin{equation}
\rho(x',x;\sigma',\sigma)=\sum_{m,n=1}^{N}\rho_{m,n}(x',x)S_{m,n}(\sigma',\sigma)\,.\label{OBDMspincharge}
\end{equation}
Equation (\ref{OBDMspincharge}) takes a kind of ``spin-charge" separated form:
The spatial part  
\begin{equation}
\begin{split}
\rho_{m,n}(x',x)=&(-1)^{n-m}N!\int_{\Gamma_{m,n}}dx_{1}...dx_{N-1} \\
&\cdot\varphi^{*}(x_{1},...,x_{N-1},x') \,\varphi(x_{1},...,x_{N-1},x)\,,\label{SOBDM}
\end{split}
\end{equation}
depends only on the charge state $\varphi$.
The information on the spin degrees of freedom is carried by the spin correlation function 
\begin{equation}
S_{m,n}(\sigma',\sigma)=(\pm1)^{m-n}\braket{\chi|S_m^{\sigma',\sigma}(m...n)|\chi}\,, \label{Smn}
\end{equation}
(again, $\pm1$ for bosonic and fermionic gases, respectively) where $S_m^{\sigma',\sigma}$ is a local SU($N$) generator ($S^{\sigma',\sigma}\ket{\sigma}=\ket{\sigma'}$) on site $m$, and $(m...n)$ is a loop permutation operator that permutes $m\rightarrow m+1,m+1\rightarrow m+2,...,n-1\rightarrow n,n\rightarrow m$. In the above, we have assumed that $m\le n$. The case with $m\ge n$ can be obtained using the identity $\rho_{m,n}(x',x) =\rho_{n,m}(x,x')$ and $S_{m,n}(\sigma',\sigma)=  S_{n,m}(\sigma,\sigma')$.

The difficulty of evaluating the OBDM lies in the fact that Eq.~(\ref{SOBDM}) involves an ($N-1$)-dimensional integral. With sophisticated numerical techniques, one may be able to carry out such an integral up to $\sim N=20$ \cite{Deuretzbacher2016}. Here we develop a new method to evaluate $\rho_{m,n}(x',x)$, which relies on its discrete Fourier transform given by:
\begin{equation}
\rho_{m,n}(x',x) = N^{-2} \sum_{\kappa, \kappa'} \rho^{\kappa',\kappa}(x',x) \,e^{i\pi\kappa'm} \,e^{-i\pi\kappa n}\,,
\end{equation}
where $\kappa$ and $\kappa'$ take a discrete set of values $2k/N$ with $N$ consecutive integers $k$, and
\begin{widetext}
\begin{equation}
\rho^{\kappa',\kappa}(x',x)=N\int dx_{1}...dx_{N-1}\prod_{j=1}^{N-1}A^{\kappa'*}(x_{j}-x')A^{\kappa}(x_{j}-x) \,\varphi^{*}(x_{1}...x_{N-1},x') \,\varphi(x_{1}...x_{N-1},x) \,,\label{rhok1k2}
\end{equation}
\end{widetext}
where $A^{\kappa}(x_{i}-x_{j})\equiv e^{i\pi(1-\kappa)\theta(x_{i}-x_{j})}$, with $\theta(x)$ being the Heaviside step function. Remarkably,
\begin{equation}
\Psi^\kappa (x_1,...,x_N) =\left[ \prod_{i<j} A^{\kappa}(x_{j}-x_i) \right] \,\varphi(x_1,...,x_N)\,,\label{psikappa}
\end{equation}
is the wave function of $N$ hard-core spinless anyons~\cite{Zhu1996, Girardeau2006} with statistical parameter $\kappa$ (we use the convention in Ref.~\cite{Calabrese2007,Santachiara2007, Santachiara2008}), whose OBDM, $\rho^\kappa(x',x)\equiv \rho^{\kappa,\kappa}(x',x)$, is given exactly by Eq.~(\ref{rhok1k2}) with $\kappa'=\kappa$. The case with $\kappa=0$ and 1 correspond to the hard-core spinless bosons and the ideal spinless fermions, respectively. By defining a similar Fourier transform for the spin correlation function \[S^{\kappa',\kappa}=N^{-2}\sum_{m,n=1}^{N}S_{m,n}e^{i\pi\kappa'm}e^{-i\pi\kappa n}\,,\] we can rewrite Eq.~(\ref{OBDMspincharge}), the OBDM of a strongly interacting spinor quantum gas, as 
\begin{equation}
\rho(x',x;\sigma',\sigma)=\sum_{\kappa',\kappa}\rho^{\kappa',\kappa}(x',x)S^{\kappa',\kappa}(\sigma',\sigma) \,. \label{mainresult}
\end{equation}
There has been an extensive study of the properties of 1D hard-core spinless anyon gases \cite{Zhu1996,Girardeau2006,Calabrese2007,Santachiara2007,Santachiara2008,Kundu1999,Batchelor2006_2007,Patu2008_2010,Rigol2004-2014,Hao2008_2009,Hao2016,Marmorini2016,Papenbrock2003,Forrester2003} (and the references therein). In particular, their OBDM and momentum distributions have been calculated. We can take advantage of these results to evaluate Eq.~(\ref{mainresult}) in a very efficient way. In the following, we present two examples, one concerns a homogeneous system with translational invariance and the other a harmonically trapped system. And for both of these two cases, we consider $\varphi$ as the ground state slater determinant.

 
\section{Translational Invariant System} 

For a translational invariant system with length $L$ (periodic boundary condition is assumed), the OBDM $\rho(x',x;\sigma',\sigma)$ depends only on $y\equiv x-x'$, and Eqs.~(\ref{OBDMspincharge}) and (\ref{mainresult}) are reduced to
\begin{eqnarray}
\rho(x',x;\sigma',\sigma) &=& \sum_{r=0}^{N-1} \rho_r(y)\, S_r(\sigma',\sigma) \nonumber \\
&=& \sum_{\kappa} \rho^\kappa(y) S^\kappa (\sigma',\sigma) \,,\label{hobdm}
\end{eqnarray}
 where $r$ in the first line is understood as $n-m$, so from Eq.~(\ref{Smn}) we have $S_r(\sigma',\sigma) = (\pm1)^{r}\braket{\chi|S_m^{\sigma',\sigma}(m...m+r)|\chi}$ which is independent of $m$, and in the second line $S^{\kappa}=N^{-1}\sum_{r=0}^{N-1}S_r e^{-i\pi\kappa r}$ only depends on spin. To ensure the boundary condition, we need to impose the selection rule $(1...N)\chi=(\mp1)^{N-1}\chi$ on the spin state $\chi$ with $\mp1$ for bosonic and fermionic gases, respectively. After Fourier transform with respect to $y$, the corresponding momentum distribution for the spinor quantum gas can be obtained as
\begin{equation}
\rho_\sigma(p)=\sum_{\kappa}\rho^{\kappa}(p)\,S^{\kappa}(\sigma,\sigma) \,,\,\label{MomentumTI}
\end{equation}
where $\rho^{\kappa}(p)$ is the momentum distribution for the hard-core anyon system. Note that $\rho^\kappa$ and $S^\kappa$ are periodic in $\kappa$ with period 2. Hence we may restrict $\kappa$ in the range $[-1, 1]$.

The OBDM for the homogeneous hard-core anyon gas, $\rho^{\kappa}(y)$, has an analytic expression in the form of the Toeplitz determinant \cite{Calabrese2007,Santachiara2007, Santachiara2008}. Its momentum distribution, $\rho^{\kappa}(p)$, is investigated in Ref.~\cite{Santachiara2008}. It is shown that $\rho^{\kappa}(p)$ is peaked at $p=\kappa \hbar k_F$, where $k_F=N\pi/L$ is the Fermi momentum, for $\kappa \in (-1,1)$. Whereas for $\kappa=\pm 1$, the system becomes an ideal spinless Fermi gas whose momentum distribution is characterized by the Fermi sea. Examples of $\rho^\kappa(p)$ for $N=201$ are shown in Fig.~\ref{Fig2}(c).

\begin{figure}[h]
\includegraphics[width=8cm]{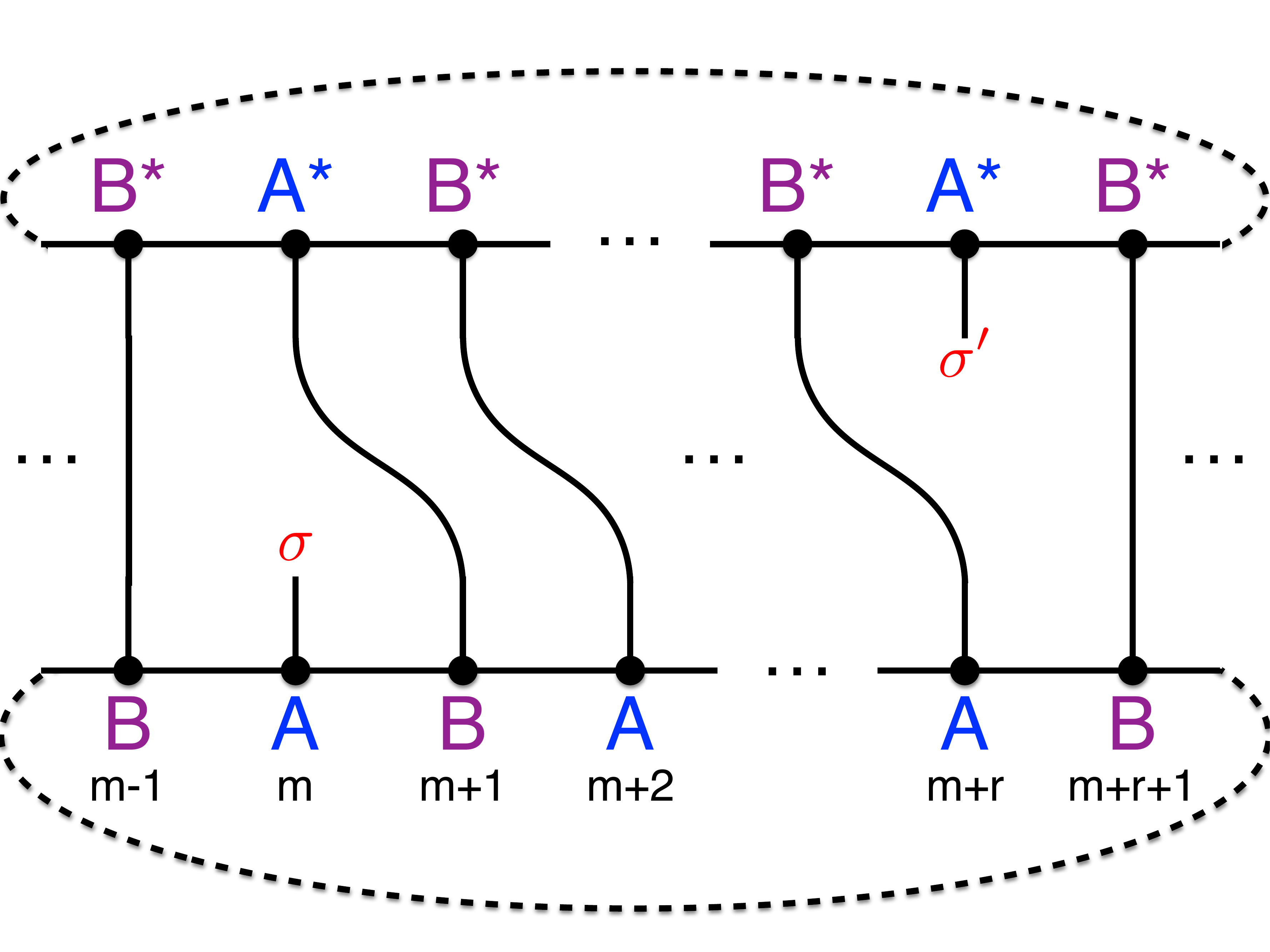}
\caption{The tensor contraction geometry for calculating $S_r(\sigma',\sigma)$ for an even $r$ case. $A$ and $B$ tensors, which are building blocks in MPS (two sites in a unit cell), are calculated using the iTEBD method. Note that for a finite periodic boundary condition system, we also need to contract the remaining tensors outside the correlation range $m$ to $m+r$. Starting from the $m^{\rm th}$ site with either $A$ tensor or $B$ tensor gives the same result.}\label{Fig1}
\end{figure}

\begin{figure}[h]
\includegraphics[width=8.9cm]{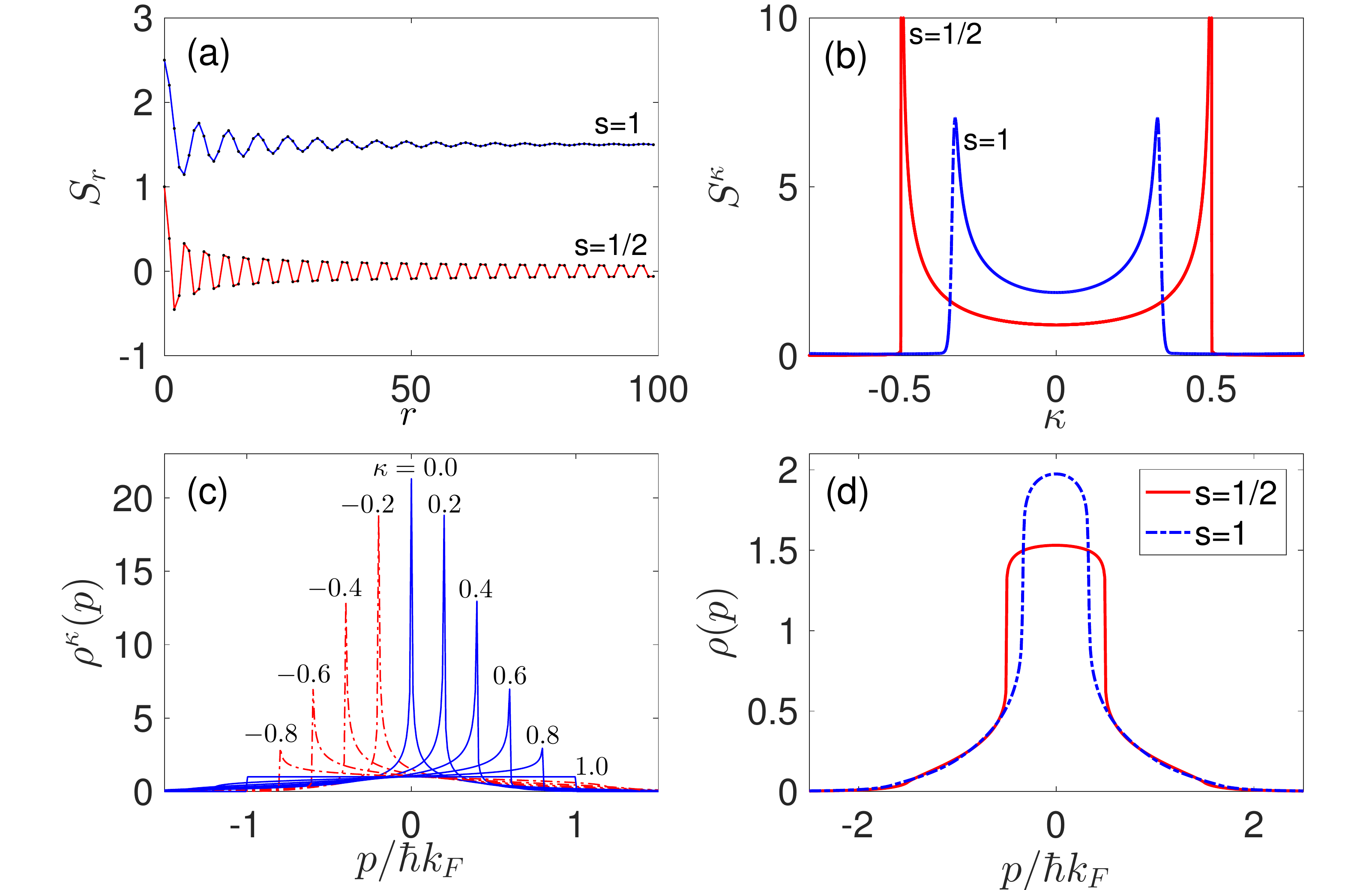}
\caption{(color online) Spin correlation function and momentum distribution of translational invariant system. {\bf (a)} $S_r$ calculated by iTEBD for an infinite chain. {\bf (b)} $S^{\kappa}$ obtained by Fourier transform of $S_r$ with $r$ up to 10000. {\bf (c)} Momentum distribution of hard-core anyon gas $\rho^{\kappa}(p)$ for $N$=201. {\bf (d)} Momentum distribution (summed over all spin components) of the spinor gases for $N$=201 particles. Note that in Eq.~(\ref{MomentumTI}), $\rho^{\kappa}(p)$ and $S^{\kappa}$ are not generally real valued as in {\bf (b)(c)}, but we can rearrange $\rho_r(y)$ and $S_r$ to make them real \cite{makereal}.}\label{Fig2}
\end{figure}

To find the OBDM and the momentum distribution of a spinor gas, all we need to do is to calculate the spin correlation functions $S_r(\sigma',\sigma)$ or $S^\kappa(\sigma',\sigma)$ and plug it into Eqs.~(\ref{hobdm}) and (\ref{MomentumTI}). For 1D system, Matrix Product State (MPS) is a representation that efficiently captures the bipartite entanglement, and many powerful methods based on this representation such as Density Matrix Renormalization Group (DMRG) and Time-Evolving Block Decimation (TEBD) have been developed to calculate the ground state and the time evolution. We calculate the ground state $S_r(\sigma',\sigma)$ using the infinite system size TEBD (iTEBD) method \cite{Vidal2007,Kjall2013}. We first calculate the $A$, $B$ tensors (two sites in one unit cell), which are building blocks in Matrix Product States(MPS), using iTEBD. Note that $S_r(\sigma',\sigma)$ is the correlation function containing a loop permutation operator $(m...m+r)$, so we use the tensor contraction geometry schematically shown in Fig.~\ref{Fig1} to calculate $S_r(\sigma',\sigma)$, and then take the Fourier transform to obtain $S^{\kappa}(\sigma',\sigma)$.

As examples, we consider a spin-1/2 and a spin-1 Fermi gases with spin-independent interaction with $N=201$. The corresponding spin-chain models in the strong interaction limit are the SU(2) and the SU(3) Sutherland models, respectively \cite{Sutherland1975}. The spin correlation functions $S_r=\sum_{\sigma}S_r(\sigma,\sigma)$ and $S^{\kappa}=\sum_{\sigma}S^{\kappa}(\sigma,\sigma)$ are plotted in Fig.~\ref{Fig2}(a) and (b), respectively. The total momentum distribution functions $\rho(p)=\sum_{\sigma}\rho_{\sigma}(p)$ for the spinor gas are shown in Fig.~\ref{Fig2}(d). The spinor quantum gas in strongly repulsive regime has been studied within the context of spin-incoherent Luttinger liquid \cite{Fiete2007}, and the ground state momentum distribution for SU(2) case has been studied in Ref.~\cite{Ogata1990,Cheianov2005,Imambekov2006}, the result in Fig.~\ref{Fig2}(d) can be compared with Fig. 3 in Ref.~\cite{Ogata1990} which is for a lattice system and for up to 32 sites with a quarter filling (note that their definition of $k_F$ differs from ours by a factor of 2). Here we want to mention that a sophisticated method developed in Ref.~\cite{Imambekov2006} can be used to efficiently calculate $\rho(p)$ for homogeneous spin-1/2 fermions, but our method is more flexible and much more general as it deals with bosonic or fermionic systems with arbitrary spin.

 
\section{Trapped System} 

For trapped systems, the OBDM is calculated using Eq.~(\ref{mainresult}), where $\rho^{\kappa',\kappa}(x',x)$ is defined with Eq.~(\ref{rhok1k2}). Unlike in the homogeneous system, we now need both the diagonal elements with $\kappa=\kappa'$ and the off-diagonal elements with $\kappa \neq \kappa'$, the latter of which can be regarded as a straightforward generalization of the OBDM of a spinless anyon gas. For the case that $\varphi$ is a slater determinant composed of single particle wave functions $\phi_j(x)$ with $j=1,2,...,N$ simply being labels, which means we can separate the coordinate $x$ as
\begin{equation}
\begin{split}
&\varphi(x_1,...,x_{N-1},x)\\
=&\frac{1}{\sqrt{N!}}\sum_P (-1)^P P(\phi_1(x_1)\phi_2(x_2)...\phi_{N-1}(x_{N-1})\phi_N(x))\\
=&\frac{1}{\sqrt{N!}}\sum_{n=1}^{N}(-1)^{N-n}\phi_{n}(x)\text{det}\left[\phi_{1,...,n-1,n+1,...,N}^{x_{1},...,x_{N-1}}\right]\,, \label{arbtrap_1}
\end{split}
\end{equation}
similarly for $\varphi^*(x_1,...,x_{N-1},x')$. We need to substitute them into Eq.~(\ref{rhok1k2}). First combine the fully symmetric direct product function $\prod_{j=1}^{N-1}A^{\kappa}(x_{j}-x)$ with $\text{det}\left[\phi_{1,...,n-1,n+1,...,N}^{x_{1},...,x_{N-1}}\right]$ to form a new determinant
\begin{equation}
\begin{split}
&\prod_{j=1}^{N-1}A^{\kappa}(x_{j}-x)\text{det}\left[\phi_{1,...,n-1,n+1,...,N}^{x_{1},...,x_{N-1}}\right]\\
=&\text{det}\left[(A^{\kappa}(x)\circ\phi)_{1,...,n-1,n+1,...,N}^{x_{1},...,x_{N-1}}\right]\,, \label{arbtrap_2}
\end{split}
\end{equation}
where $A^{\kappa}(x)\circ\phi$ means using $A^{\kappa}(t-x)\phi_k(t)$ as basis of the slater determinant. Next using the identity 
\begin{equation}
\begin{split}
&\int dx_{1}...dx_{N-1}\text{det}\left[(A^{\kappa'*}(x')\circ\phi)_{1,...,m-1,m+1,...,N}^{x_{1},...x_{N-1}}\right]\\
&\cdot\text{det}\left[(A^{\kappa}(x)\circ\phi)_{1,...,n-1,n+1,...,N}^{x_{1},...x_{N-1}}\right]\\
=&(N-1)! \,\text{det}\left[\hat{\Phi}^{(m,n)}\right]\,, \label{arbtrap_3}
\end{split}
\end{equation}
which can be easily proved, where $(m,n)$ stands for minor, which is the determinant of a matrix after deleting its $m^{\rm th}$ row and $n^{\rm th}$ column, and the matrix $\hat{\Phi}$ depends on $\kappa',\kappa,x',x$, with its elements given by
\begin{equation}
\hat{\Phi}_{k,l}^{\kappa',\kappa}(x',x)=\int^{\infty}_{-\infty} dtA^{\kappa'*}(t-x')A^{\kappa}(t-x)\phi^{*}_{k}(t)\phi_{l}(t)\,,
\end{equation}
where $\phi_{k,l}$ are single-particle wave functions and $k,l=1,2,...,N$. Finally putting Eqs.~(\ref{arbtrap_1}) $\sim$ (\ref{arbtrap_3}) together, Eq.~(\ref{rhok1k2}) can be written into a form with only minors of a determinant:
\begin{equation}
\rho^{\kappa',\kappa}(x',x)=\sum_{m,n}(-1)^{m-n}\phi^{*}_{m}(x')\phi_{n}(x)\text{det} \left[\hat{\Phi}^{(m,n)}\right] \,. \label{arbtrapped}
\end{equation}
The expression Eq.~(\ref{arbtrapped}) is much simpler than the previous formulas for OBDM as reported in  Ref.~\cite{Deuretzbacher2016}, which rely on the calculation of Taylor coefficients of matrix determinants using sophisticated methods \cite{Deuretzbacher2008,Loft2016_1,Loft2016_2,Yang2016,Deuretzbacher2016}.

\subsection{Harmonically Trapped System}
For the most experimentally relevant harmonically trapped systems, an even simpler form of the OBDM can be obtained as follows. Note that wave function $\varphi$ of a harmonically trapped spinless fermion can be written into a Vandermonde determinant form:
\begin{equation}
\begin{split}
&\varphi(x_{1},...,x_{N})=\frac{1}{\sqrt{N!}}\text{det}\left[\phi_{0,1,...,N-1}^{x_1,...,x_{N}}\right]\\
=&C_{N}^{1/2}\prod_{j=1}^{N}e^{-x_{j}^{2}/2}\prod_{1\le j<k\le N}(x_{j}-x_{k})\,, \label{Vandermonde}
\end{split}
\end{equation}
where $\phi_{0,1,...,N-1}^{x_1,...,x_{N}}$ means the slater determinant uses single particle harmonic oscillator wave functions $\phi_{k}(x)$($k=0,1,...,N-1$) as basis. And
\begin{equation}
C_{N}=\frac{2^{N(N-1)/2}}{\pi^{N/2}\left[\prod_{n=1}^{N}n!\right]}
\end{equation}
is a normalization constant. This leads to 
\begin{widetext}
	\begin{equation}
	\varphi(x_{1},...,x_{N-1},x)
	=C_{N}^{1/2}e^{-x^{2}/2}\prod_{j=1}^{N-1}(x_{j}-x)\prod_{j=1}^{N-1}e^{-x_{j}^{2}/2}\prod_{1\le j<k\le N-1}(x_{j}-x_{k})\,, \label{harmonictrap_1}
	\end{equation}
	which after substituting into Eq.~(\ref{rhok1k2}), and using the $N-1$ version of Eq.~(\ref{Vandermonde}), we have 
	\begin{eqnarray*}
	\rho^{\kappa^{'},\kappa}(x',x)
	& =& NC_{N}e^{-\frac{x'^{2}+x^{2}}{2}}\int dx_{1}...dx_{N-1}\prod_{j=1}^{N-1}A^{\kappa'*}(x_{j}-x')(x_{j}-x')A^{\kappa}(x_{j}-x)(x_{j}-x)\prod_{j=1}^{N-1}e^{-x_{j}^{2}} \!\!\! \prod_{1\le j<k\le N-1} \!\!\! \!\!\!(x_{j}-x_{k})^{2}\\
&
	=& \frac{NC_{N}e^{-\frac{x'^{2}+x^{2}}{2}}}{C_{N-1}(N-1)!}\int dx_{1}...dx_{N-1}\prod_{j=1}^{N-1}A^{\kappa'*}(x_{j}-x')(x_{j}-x')A^{\kappa}(x_{j}-x)(x_{j}-x)\left(\text{det}\left[\phi_{0,1,...,N-2}^{x_1,...,x_{N-1}})\right]\right)^{2}\,.
	\end{eqnarray*}
\end{widetext} 
Now by using a similar procedure as in arbitrary trapping potential case that leads to Eq.~(\ref{arbtrapped}), we can combine the product of  $\prod_{j=1}^{N-1}A^{\kappa'*}(x_{j}-x')(x_{j}-x')A^{\kappa}(x_{j}-x)(x_{j}-x)$ into the square of a determinant to form a square of a new determinant, and then carry out the $(N-1)$-dimensional integral. Finally we arrive at the following: 
\begin{equation}
\rho^{\kappa',\kappa}(x',x)=\frac{e^{-(x'^{2}+x^{2})/2}}{\pi^{1/2}}\text{det}\left(\hat{B}\right)\,,
\end{equation}
where the elements of the matrix $\hat{B}$ are
\begin{equation}
\begin{split}
&\hat{B}_{k,l}^{\kappa',\kappa}(x',x) = 2/\sqrt{(k+1)(l+1)} \\
& \times \int^{\infty}_{-\infty} \!\!dt\,A^{\kappa'*}(t-x')A^{\kappa}(t-x)(t-x')(t-x) {\phi}^{*}_{k}(t) {\phi}_{l}(t),
\end{split}
\end{equation}
where $\phi_{k,l}$ are single particle eigen wave functions of harmonic oscillator, and $k,l=0,1,...,N-2$.

The OBDM of a hamonically trapped hard-core spinless anyon gas $\rho^{\kappa}(x',x)=\rho^{\kappa,\kappa}(x',x)$ have been investigated previously \cite{Hao2016, Marmorini2016} (for hard-core spinless Bose gas, see Ref.~\cite{Papenbrock2003, Forrester2003}). 

\subsection{Impurity in a Tonks-Girardeau Gas}
As a concrete example, we consider a
recent experiment \cite{Meinert2016} where Bloch oscillation of a single impurity atom moving in the background of a strongly interacting spinless Bose gas (i.e., the Tonks-Girardeau gas) is observed. Here, we explain this phenomenon using the strong coupling ansatz with the spin-chain model theory, which is a different perspective from previous theoretical studies \cite{Gangardt2009,Schecter2012,Gamayun2014,Schecter2016}. 

We model the system as a spin-1/2 Bose gas with atomic mass $m$, confined in a harmonic trap with trapping potential $\omega$, with one spin-$\downarrow$ atom as the impurity and ($N-1$) spin-$\uparrow$ atoms as the background. Strong repulsive interaction exists between the background atoms, and also between the background and the impurity atoms. In this strong interaction regime, we can write down a spin-chain model. However, for this particular system with one single impurity, we can model the dynamics of the impurity atom as if it hops on an effective lattice under the influence of a constant force $F$. It can be easily proved that the Hilbert space of this one atom hopping model and that of the spin-chain model governed by Hamiltonian~(\ref{Hsc}) with one spin impurity are equivalent. The Hamiltonian of the one atom hopping model takes the following form (setting $\hbar=m=\omega=1$)
\begin{equation}
\begin{split}
H_{sc}=&-\frac{\pi}{\sqrt{2N}\gamma_{i}}\sum_{j=1}^{N-1}C_{j}\left[c_{j}^{\dagger}c_{j+1}+h.c.\right]\\
&+\left[\frac{1}{\pi}\sqrt{2N}\right]^{3}{\cal F}\sum_{j=1}^{N-1}D_{j}n_{j} \,,\label{impHsc}
\end{split}
\end{equation}  
where $\gamma_{i}=mg_{i}/\hbar^{2}n_{1D}$ is the dimensionless interaction constant, with $n_{1D}=\sqrt{2N}/\pi$ the density at trap center, and $g_{i}$ the contact interaction strength between the impurity and the background atoms \cite{note}. $H_{sc}$ is a single-atom Hamiltonian. $c^{\dagger}_j$ and $c_j$ are creation and annihilation operators for this single atom, and $n_j=c^{\dagger}_jc_j$ are local density operators. The first line of (\ref{impHsc}) represents the kinetic term and the second line the force term. The coupling coefficients $C_j$ can be calculated using a special local density approximation method \cite{LDAonCi}. The force on the impurity is modeled as a magnetic gradient and represented by the second line in (\ref{impHsc}) where ${\cal F}=mF/\hbar^{2}n_{1D}^{3}$ and $D_j=C_{j-1}-C_{j}$ (assuming $C_{0}=C_{N}=0$) \cite{Yang2016}.

\begin{figure}[h]
\includegraphics[width=9cm]{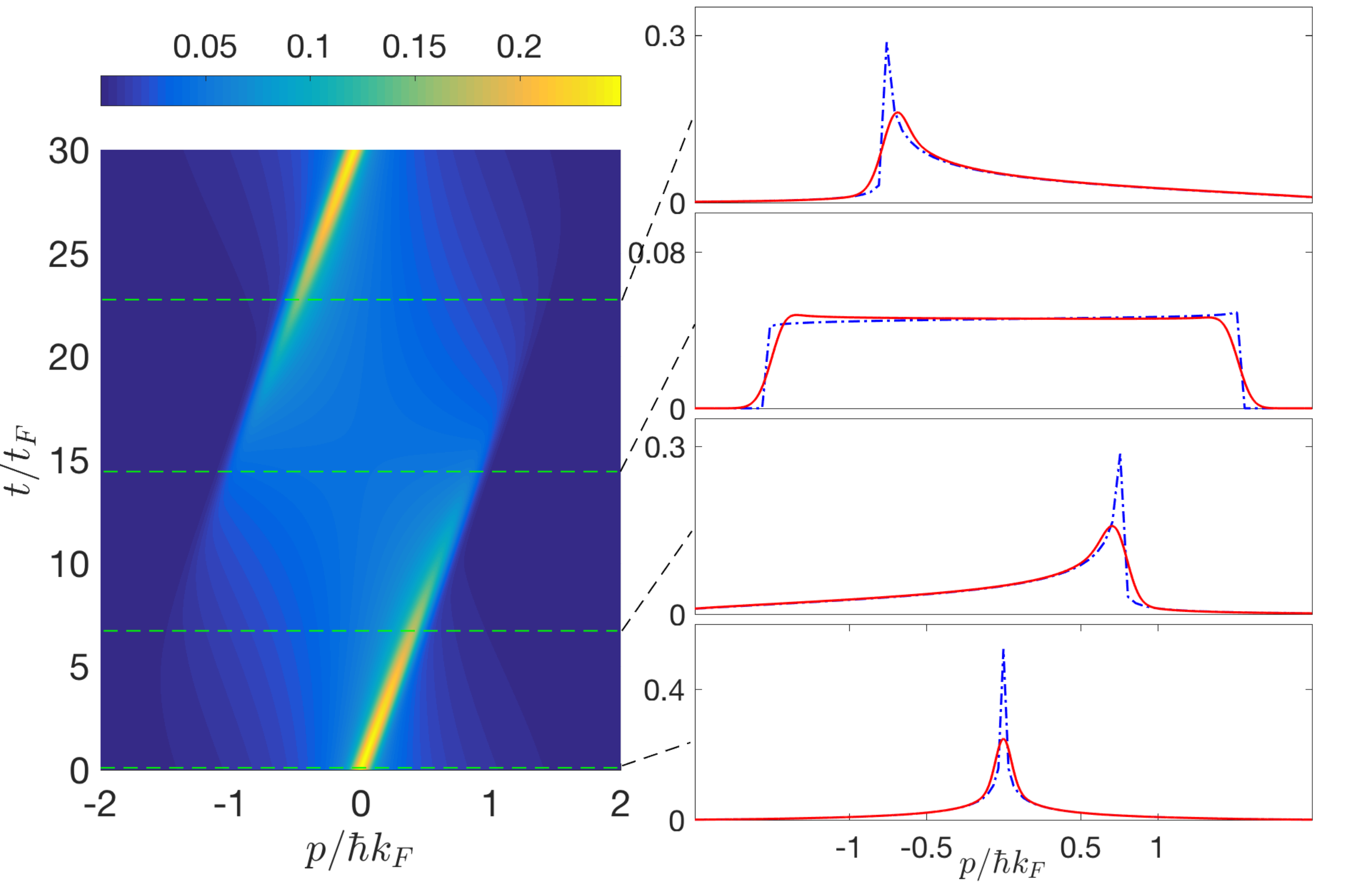}
\caption{(color online) Left panel: evolution of the momentum distribution of the impurity atom. Here we take $N=60$, $\gamma_i=12$, and ${\cal F}=1$. $t_{F}=\hbar/E_{F}=1/N$, and $k_{F}=n_{1D}\pi={\sqrt{2N}}$ is the Fermi momentum. All quantities are expressed in the dimensionless unit system defined by $\hbar=m=\omega=1$. Right panel: the solid lines replot the momentum distribution of the impurity atom from the left panel at four different times; the dash-dotted line is the momentum distribution of a homogeneous hard-core anyon gas, $\rho^\kappa(p)/N$, with statistical parameter $\kappa=Ft/\hbar k_F$. The anyon gas consists $N$ particles confined in a region with length $L$ (periodic boundary condition is assumed) such that its density is given by $N/L=n_{1D}$. }\label{Fig3}
\end{figure}

We take the initial spin state to be the ground state of Hamiltonian (\ref{impHsc}) in the absence of the force term, which subsequently evolves in time under the full Hamiltonian (\ref{impHsc}). With the instantaneous spin state obtained by solving the Schr\"{o}dinger equation \cite{note1}, and using the method outlined above, we can calculate the momentum distribution of the impurity spin which we plot on the left panel of Fig.~\ref{Fig3}. The initial momentum distribution is peaked at $p=0$ as expected. This peak moves towards the Fermi point $\hbar k_F$ as the impurity is under the influence of the force. When the peak reaches $\hbar k_F$, it disappears and re-emerges at the other Fermi point $-\hbar k_F$. Thus the impurity atom carries out the Bloch oscillation. Our calculation agrees qualitatively with the experiment of Ref.~\cite{Meinert2016}.

Another interesting aspect of this experiment is that the measured momentum distribution of the impurity atom is approximately the momentum distribution of a hard-core anyon gas with a time-dependent statistical parameter $\kappa$. To see this, let us ignore the trapping potential, which is not essential for the Bloch oscillation dynamics, and assume that the system is homogeneous for simplicity. In this case, the initial spin state has exactly zero momentum with $S^{\kappa}=\delta_{\kappa,0}/N$. If $\gamma_i$ is sufficiently large, we may ignore the hopping term, i.e., the first line of Hamiltonian (\ref{impHsc}). Under this approximation, the spin correlation function evolves simply as $S^\kappa(t) =\delta_{\kappa,Ft/\hbar k_F}/N$. According to Eq.~(\ref{MomentumTI}), the momentum distribution of the impurity atom at time $t$ is thus given by 
\[\rho(p,t)=\frac{1}{N}\rho^{Ft/\hbar k_F}(p)\,,\] 
which is exactly the momentum distribution of a hard-core anyon gas with a time-dependent statistical parameter $\kappa=Ft/\hbar k_F$. On the right panel of Fig.~\ref{Fig3}, we replotted the momentum distribution of the impurity atom obtained above at several different times (solid lines), and compared them with the momentum distribution of a homogeneous hard-core anyon gas with its density given by $n_{1D}$, particle number $N$, and $\kappa=Ft/\hbar k_F$ (dash-dotted lines). Good qualitative agreement can be seen. The main difference is that the distribution of the trapped impurity atom has a rounded peak, which can be mainly attributed to the effect of the trapping potential. 


\section{Conclusion}

In conclusion, we have shown that the OBDM of a 1D strongly interacting spinor quantum gas and that of the spinless hard-core anyons are related to each other by a Fourier transform. This allows us to write down the OBDM of a strongly interacting spinor gas in a simple form as represented by Eq.~(\ref{mainresult}), which is valid for systems with arbitrary spin and arbitrary trapping potentials. For certain special cases, such as homogeneous or harmonically trapped systems, the OBDM of the anyon gas possess closed forms, which allows us to efficiently calculate the OBDM of a spinor gas with much larger particle numbers than what was previously possible.
The OBDM is essentially a nonlocal correlation function, with which one can easily calculate the momentum distribution of the system, as illustrated in this work. Momentum distributions of cold atoms are routinely measured in experiment. They provide crucial information about the quantum states of the system. Our work therefore not only provides a powerful method to calculate these quantities very efficiently, but will also shed new light onto 1D quantum many-body systems in the strong interaction limit. 
 
\begin{acknowledgments}
We would like to thank Hanns-Christoph N\"{a}gerl for providing us with their preprint of Ref.~\cite{Meinert2016} before it was made public. Their experiment motivated us to study the momentum distribution of a strongly interacting 1D system. We also thank Xiwen Guan and Matthew S. Foster for their helpful and inspiring discussions, and Jiyao Chen for detailed discussion on iTEBD. This research is supported by the US NSF and the Welch Foundation (Grant No. C-1669).
\end{acknowledgments}



\begin{thebibliography}{99}
\bibitem{makereal} For a system with periodic boundary condition, for odd (even) $N$, $\rho^{\kappa}(p)$ and $S^{\kappa}$ are only purely real (imaginary) when $\kappa=2m/(N-1)$ with integer $m$. In Eq.~(\ref{MomentumTI}), we use $\kappa=2m/N$. But we can redefine $\rho^{\kappa}(p)$ and $S^{\kappa}$ to make them real by rearranging $\rho_r(y)$ and $S_r$ to make them mirror symmetric about $y=0$ and $r=0$. For small number of particles, this leads to small oscillations in $\rho^{\kappa}(p)$ and $S^{\kappa}$(no effect on $\rho(p)$). But these oscillations tend to vanish in the thermodynamic limit.
\bibitem{note} The exchange between two background atoms only contributes a constant energy shift, which can be ignored. This is why the interaction strength between the background atoms do not appear in the Hamiltonian (\ref{impHsc}).
\end{thebibliography}
\end{document}